\documentclass[floatfix,footinbib,reprint,twocolumn,prx,aps,a4paper,preprintnumbers,amsmath,amssymb,longbibliography]{revtex4-1}

\usepackage{graphicx} %
\usepackage{bbold} 
\usepackage{bm}
\usepackage{color}
\usepackage[colorlinks=true,linkcolor=blue,citecolor=blue,urlcolor=blue]{hyperref}

\date{\today}

\begin{document}
\title{\bf Caloric properties of materials near a tricritical point}
\author{Eduardo Mendive-Tapia}
\email{e.mendive.tapia@ub.edu}
\thanks{Corresponding author}
\affiliation{Departament de Física de la Matèria Condensada, Facultat de Física, Universitat de Barcelona, Martí i
Franquès 1, E-08028 Barcelona, Catalonia}

\author{Antoni Planes}
\email{antoniplanes@ub.edu}
\thanks{Corresponding author}
\affiliation{Departament de Física de la Matèria Condensada, Facultat de Física, Universitat de Barcelona, Martí i
Franquès 1, E-08028 Barcelona, Catalonia}

\begin{abstract}
We present a mean-field study of caloric effects driven by primary and secondary fields near tricritical points,
which are fields thermodynamically conjugated to the main and secondary order parameters, respectively. General features, such as critical exponents and their crossover from critical to tricritical behaviours, are studied by means of a generic free energy Landau expansion. To deal with specific materials, we propose a model that combines the Blume-Emery-Griffiths prototype to study tricritical points with the Bean-Rodbell approach to include magnetovolume effects.
In this model the primary field is the magnetic field, while chemical and mechanical pressures are secondary fields.
In spite of the scarcity of experimental data, we have shown that results for the La(Fe$_{x}$Si$_{1-x}$)$_{13}$ and MnSi compounds are in good agreement with our predictions. 
We expect that our results will motivate and guide new experimental research aiming at optimizing caloric materials.
\end{abstract}

\maketitle

\section{Introduction}
\label{intro}

Caloric properties of materials are determined by their reversible thermal response to an external field. When the field is applied/removed keeping the temperature constant the response can be quantified by a change of entropy, while it is measured by a change of temperature when the field is applied or removed in adiabatic conditions~\cite{Manosa2013,Moya2014}. These changes are expected to be very large in the vicinity of a phase transition and, particularly, near a first-order transition due to the contribution of the latent heat to the thermal response. Nevertheless, first-order transitions take place under non-equilibrium conditions that are intrinsically associated with the nucleation of the new phase and cause a reduction of the expected reversible contribution to the caloric effect. This is in contrast with continuous transitions, which occur in nearly equilibrium conditions in general. Materials undergoing a first-order transition are expected to have enhanced caloric properties. However, in practice the need of larger fields to overcome hysteresis effects can make materials that undergo a continuous transition to show a comparable  reversible thermal response for the same applied field. In fact, it has been suggested that the competition between latent heat and hysteresis should be optimized near the end point of a first-order transition line, where the nature of the transition changes from first-order to continuous. This boundary defines a tricritical point, whose search in materials represents a great interest for the development of optimal caloric responses~\cite{PhysRevLett.104.247202,Sandemann2012}.

Alongside a tricritical point there exists a secondary parameter that is coupled to the main order-parameters and can be controlled by an external secondary field. This opens the possibility of inducing caloric effects not only by applying the field thermodynamically conjugated to the order parameter, but also by inducing the thermal change via fields thermodynamically conjugated to secondary parameters. An example of such a situation is the famous La(Fe$_x$Si$_{1-x}$)$_{13}$ magnetocaloric compound~\cite{PhysRevB.65.014410,PhysRevB.67.104416}. While for $x$ larger than 0.89 (Fe-rich content) the ground state is antiferromagnetic, this material becomes ferromagnetic for $x$ lower than 0.89~\cite{PhysRevB.65.014410}. In a relatively narrow range, close to the composition that separates the ferromagnetic and antiferromagnetic ground states, the transition from the paramagnetic phase to the ferromagnetic phase is first-order and a tricritical point is reached for $x = 0.86$. The transition is continuous for lower Fe-contents. Since the first-order transition temperature of this compound can be controlled by an applied magnetic field, mechanical pressure (thanks to a strong magneto-volumic interplay) and chemical pressure (composition), the tricritical point can be approached by applying a magnetic field (thermodynamically conjugated to the order parameter), decreasing the mechanical pressure that consequently controls the volume change, or the composition. It is worthwhile noting that, as expected, this compound displays both magnetocaloric and barocaloric effects~\cite{Lyubina2008}, which interestingly seem to be optimally close to the tricritical point.

The equiatomic FeRh compound shows a similar behaviour. It undergoes a metamagnetic phase transition from a low temperature antiferromanetic structure to a high temperature ferromagnetic phase, which is accompanied by a large volume change~\cite{JBMcKinnon_1970,PhysRevB.77.184401,PhysRevB.89.054427,Mendive2015}. Caloric effects can be consequently induced by magnetic field and pressure~\cite{Stern-Taulats2017}. While a tricritical point also exists, the caloric behaviour close to this point has not been studied since it requires application of a very high applied pressure of about 6 GPa~\cite{10.1063/1.1658533,https://doi.org/10.1002/pssb.2220780136}.
The heavy rare earth elements also show tricritical points approached by both a magnetic field (primary field)~\cite{HERZ1978273,PhysRevB.71.184410,PhysRevLett.118.197202,Zverev_2015} and uniaxial pressure (secondary field)~\cite{PhysRevB.84.132401,PhysRevB.104.174435} when these are applied close to their first-order helimagnetic-ferromagnetic phase transitions.
We highlight that the similar shape of the temperature-pressure/stress phase diagrams of FeRh and Tb is also shown by the magnetically frustrated Mn$_3$Pt cubic compound~\cite{10.1063/1.1658533,https://doi.org/10.1002/pssb.2220780136,PhysRevB.104.174435,doi:10.1143/JPSJ.56.4532}. In the three diagrams the suppression of a low-temperature magnetic state in favour of a high-temperature phase coincides at the tricritical point, which is achieved by smaller uniaxial tension $\sigma=0.6$MPa~\cite{PhysRevB.104.174435} and hydrostatic pressure $p=0.3$GPa~\cite{doi:10.1143/JPSJ.56.4532,PhysRevB.99.144424} for Tb and Mn$_3$Pt, respectively.
A tricritical point reached by uniaxial strain has been also predicted from first principles in Mn-based antiperovskite materials~\cite{PhysRevB.95.184438}, whose exploitation to remove hysteresis effects in a new sort of cooling cycle has been proposed~\cite{10.1063/5.0003243}.
However, corresponding measurements of caloric effects around their tricritical points in all these materials are still unavailable.

Despite the fact that tricritical points have been observed in numerous materials, studies of the behaviour of the caloric response close to them seem scarce. Some interesting data have been reported for the MnSi compound~\cite{Samatham2017}. This magnetic material undergoes a phase transition from a paramagnetic high temperature phase  to a complex low temperature structure controlled by the competition between the applied magnetic field and the magnetocrystalline anisotropy. As a consequence of this competition, a conical skirmion lattice occurs at low temperature~\cite{doi:10.1126/science.1166767,Bauer2013}. The interesting aspect is that a tricritical point can be approached by controlling the magnetic field, which is the primary field thermodynamically conjugated to the order parameter.
Another interesting system is the hybrid organic-inorganic (CH$_3$)2NH$_2$Mg(HCOO)$_3$ perovskite that undergoes an order-disorder phase transition~\cite{10.1063/1.5049116}. In this material the caloric response has been studied close to the tricritical point, which has been approached by controlling hydrostatic pressure, the latter being a secondary field not thermodynamically conjugated to the main order parameter. The study of these two systems confirms that the caloric response shows different behaviours as the tricritical point is approached by either a primary or a secondary field. 

The goal of this work is to present a careful study of caloric effects induced close to tricritical points by either a field thermodynamically conjugated to the order parameter or by a field conjugated to a secondary parameter. The paper is organized as follows. In Sect.\ \ref{critical_exp} the general features near and at tricritical points are discussed within the frame of a Landau theory. In Sect.\ \ref{BEG} a model for a diluted ferromagnet under pressure is presented, and in Sect.\ \ref{experiments} we compare our theoretical findings with available experiments. We finalize with our conclusions in Sect.\ \ref{conclusions}.

\section{Critical and tricritrical behaviour: general features}
\label{critical_exp}

It is not uncommon to observe systems containing strongly coupled primary and secondary order parameters such that changes of one of them have repercussions on the other one. Therefore, in these systems one order parameter can be manipulated by a field thermodynamically conjugated to the other order parameter.
Here we will assume a generic system that undergoes a phase transition described by an order parameter $\phi$ which is coupled to a secondary parameter $x$. Examples may be an antiferromagnet subjected to a magnetic field, where $\phi$ would be an staggered magnetization while $x$ should be identified with the magnetization controlled by the magnetic field, or a diluted ferromagnet where the fraction of non-magnetic impurities plays the role of $x$. In the latter case the field thermodynamically conjugated to $x$ is a chemical potential.

For the sake of simplicity, we will assume that both $\phi$ and $x$ are scalars and that the system considered is invariant under change of sign of the order parameter. A Landau free energy density function adequate to describe this system is assumed to be of the general form  
\begin{equation}
\begin{split}
    \mathcal{F}(\phi,\Lambda, T)
    =\,\, & 
    a_0(\Lambda, T) + \frac{1}{2}a_2(\Lambda, T)\phi^2 \\
    & + \frac{1}{4}a_4(\Lambda, T)\phi^4 + \frac{1}{6}a_6(\Lambda, T)\phi^6,
\label{EQ_Flandau}
\end{split}
\end{equation}
where $\Lambda$ is the secondary field conjugated to the secondary parameter $x$. That is, $x$ and $\Lambda$ must be related as
\begin{equation}
    x = -\left( \frac{\partial \mathcal{F}}{\partial \Lambda} \right)_T.
    \label{x-M}
\end{equation}
Note that an expansion in powers of $\phi$ up to sixth order suffices since our interest focuses on the temperature and external field dependence of thermodynamic properties in systems with a well defined primary order parameter. The temperature-dependent coefficients are also functions of $\Lambda$ owing to the coupling between $\phi$ and $x$. We will assume that $a_2 \propto [T - T_c(\Lambda)]$ and that $a_6>0$ around the parameter space of interest, i.e.\ near a tricritical point. 

The corresponding equation of state for $\phi$ is given by
\begin{equation}
    B = \left(\frac{\partial\mathcal{F}}{\partial \phi}\right)_T = \left[
    a_2(\Lambda, T) + a_4(\Lambda, T)\phi^2 + a_6(\Lambda, T)\phi^4
    \right]\phi,
\label{EQ_derphi}
\end{equation}
where $B$ is a field thermodynamically conjugated to $\phi$. In this model a line of critical points occurs for $B = 0$ and $a_4 > 0$, which is given by the condition:
\begin{equation}
    a_2(\Lambda, T_\text{c}) = 0.
\label{EQ_Tcr}
\end{equation}

It is broadly known and follows from a simple mathematical analysis of Eq.\ (\ref{EQ_Flandau}) that the character of such a phase transition becomes discontinuous (first-order) when $a_4(\Lambda, T_c) < 0$. 
Hence, a tricritical point exists defined by the condition 
\begin{equation}
a_2(\Lambda_t, T_\text{t}) = a_4(\Lambda_t, T_\text{t}) = 0
.
\label{EQ_tricritical}
\end{equation}
It is straightforward to show that the critical exponent of the order parameter is the usual mean-field exponent $\beta = 1/2$ when $a_4 > 0$. However, near the tricritical point we must consider two different situations depending on whether $a_2$ approaches zero faster than $a_4$ (critical case) or $a_4$ approaches zero faster than $a_2$ (tricritical case):
    \begin{equation}
    \begin{split}
        &\text{Critical case:} \\ 
        & \lim_{T\rightarrow T_\text{t}^{-}} a_2(\Lambda, T) \ll \lim_{T\rightarrow T_\text{t}^{-}} a_4(\Lambda, T)\\ 
        & \Rightarrow \phi = \left(-\frac{a_2}{a_4}\right)^{1/2} \sim (T_\text{t}-T)^{1/2} \Rightarrow \beta =\frac{1}{2}
    \label{EQ_conda}
    \end{split}
    \end{equation}
    
    \begin{equation}
    \begin{split}
        & \text{Tricritical case:} \\
        & \lim_{T\rightarrow T_\text{t}^{-}} a_2(\Lambda, T) \gg \lim_{T\rightarrow T_\text{t}^{-}} a_4(\Lambda, T)\\
        & \Rightarrow \phi = \left(-\frac{a_2}{a_6}\right)^{1/4} \sim (T_\text{t}-T)^{1/4} \Rightarrow \beta_t=\frac{1}{4}
    \label{EQ_condb}
    \end{split}
    \end{equation}

In order to determine the critical behaviour of the secondary parameter $x$ we consider Eq.\ (\ref{x-M}) and write
%
\begin{equation}
\begin{split}
    - x =\,\, &
    \frac{\partial a_0}{\partial \Lambda}
    +\frac{1}{2}\frac{\partial a_2}{\partial \Lambda}\phi^2
    +\frac{1}{4}\frac{\partial a_4}{\partial \Lambda}\phi^4
    +\frac{1}{6}\frac{\partial a_6}{\partial \Lambda}\phi^6 \\
    & +\left(a_2\phi+a_4\phi^3+a_6\phi^5\right)\frac{\partial\phi}{\partial \Lambda}.
\label{EQ_derx}
\end{split}
\end{equation}
Taking into account Eq.\ (\ref{EQ_derphi}) one obtains that $a_2\phi+a_4\phi^3+a_6\phi^5 = 0$ when $B=0$. On the other hand,  the coefficients $a_n(\Lambda,T)$ can be generally expressed as a Taylor series of two variables containing both even and odd terms in $T$ and $\Lambda$,
\begin{equation}
\begin{split}
a_n(\Lambda,T) =\,\, & a_n(0,T_\text{t}) + \left.\frac{\partial a_n(\Lambda,T)}{\partial \Lambda}\right|_{\Lambda=0, T=T_\text{t}}\Lambda \\
& + \left.\frac{\partial a_n(\Lambda,T)}{\partial T}\right|_{\Lambda=0, T=T_\text{t}}(T-T_\text{t}) + \cdots
.
\label{EQ_anseries}
\end{split}
\end{equation}
Conveniently grouping terms in Eq.\ (\ref{EQ_anseries}) one can find that the lowest non-zero power in the temperature difference $T_\text{c}-T$ for the first derivative of $a_n$ with respect to $\Lambda$ is linear:
\begin{equation}
\begin{split}
   \frac{\partial a_n}{\partial \Lambda} \sim & 
    \left.\frac{\partial a_n(\Lambda,T)}{\partial \Lambda}\right|_{\Lambda=0, T=T_\text{t}} \\
    & +\frac{1}{2}\left.\frac{\partial^2 a_n(\Lambda,T)}{\partial \Lambda\partial T}\right|_{\Lambda=0, T=T_\text{t}}(T-T_\text{t}).
\label{EQ_deriva0_phi}
\end{split}
\end{equation}
The first and second leading contributions in the right hand side of Eq.\ (\ref{EQ_derx}) therefore result in the following lowest possible orders
\begin{equation}
    (x-x_c) \sim A_1(T_\text{t}-T) + A_2 \phi^2
    ,
\label{EQ_xbeta}
\end{equation}
where $A_1$ and $A_2$ are constants. $x_c$ is the value of $x$ for $B=0$ and the initial value of the applied field $\Lambda=\Lambda_c$, as well as $T=T_\text{t}$.
From Eqs.\ (\ref{EQ_conda}) to (\ref{EQ_xbeta}) one can thus determine the critical behaviour of the secondary parameter close to the tricritical point. As for $\phi$, the same two cases must be considered
\begin{equation}
         \text{Critical case:}\; 
         x-x_c \sim (T_\text{t}-T) \Rightarrow \tilde{\beta}=1
    \label{EQ_condax}
    \end{equation}
    \begin{equation}
    \begin{split}
        & \text{Tricritical case:} \\
           & x-x_t \sim
            \begin{cases}
               (T_\text{t}-T)^{1/2} \Rightarrow \tilde{\beta}_t=1/2 \;\; \text{if $T<T_\text{t}$} \\
                (T_\text{t}-T) \Rightarrow \tilde{\beta}_t=1 \;\; \text{if $T>T_\text{t}$} 
            \end{cases}
    \label{EQ_condbx}        
    \end{split}
    \end{equation}

We are now interested in studying the caloric effect close to the tricritical point. Due to the interplay between $\phi$ and $x$ the caloric effect can be induced by either application of the field $B$ conjugated to $\phi$ or the field conjugated to $x$.
These effects can be obtained by firstly observing that the entropy can be expanded in another power series
\begin{equation}
    S = S_0 + \frac{1}{2} S_2\phi^2  + \frac{1}{4} S_4\phi^4 + \cdots
\label{EQ_Sphi}
\end{equation}
Applying an external field at $T=T_\text{t}$ thus yields an isothermal entropy given as
\begin{equation}
    \Delta S = S(\phi\neq0) - S(\phi=0) \sim \frac{1}{2} S_2\phi^2
    .
\label{EQ_DSphi}
\end{equation}
From Eqs.\ (\ref{EQ_derphi}), (\ref{EQ_Tcr}), and (\ref{EQ_DSphi}) the well-known critical exponents for the entropy change with respect to the primary external field $B$ follow

\begin{itemize}
\item[] Critical case: \vspace{2mm} \\ 
    $B \approx 
    a_4(\Lambda=0, T_\text{t})\phi^3 + a_6(\Lambda=0, T_\text{t})\phi^5 \\
    \Rightarrow \phi\sim H^{1/\delta} \;\; \text{with} \; \delta = 3$, 
    which leads to 
    \begin{equation}
    \Delta S \sim B^{n} \;\; \text{with} \;\; n = 2/3.
    \label{EQ_condagamma}
    \end{equation}

\item[] Tricritical case: \vspace{2mm} \\
    $B \approx 
    a_6(\Lambda=0, T_\text{t})\phi^5 \Rightarrow \phi\sim B^{1/\delta_t} \;\; \text{with} \;\; \delta_t = 5$,
    which leads to,
    \begin{equation}
    \Delta S \sim B^{n_t} \;\; \text{with} \;\; n_t = 2/5.
    \label{EQ_condbgamma}
    \end{equation}
    
\end{itemize}
The obtained critical and tricritical exponents are sufficient for their comparison with experimental data despite their mean-field nature.
It is worth noting that they are consistent with the general expression $\Delta S \sim B^{[\beta(1 + \delta) -1]/\beta \delta}$ \cite{J.PhysC.2.1647}. It is insightful to remark that taking into account that the field-induced adiabatic temperature change goes as $\Delta T \sim B^{1/\beta \delta}$, the exponents of $\Delta S$ and $\Delta T$ must be the same in a mean-field treatment for the critical case since in this situation $\beta(1 + \delta) -1 = 1$. Instead, in the tricritical case the exponent of $\Delta T$ should be $2n_t$ = 4/5.

If we set $B=0$ and apply the secondary external field $\Lambda$ instead, Eq.\ (\ref{EQ_derphi}) reads as
\begin{equation}
    0 = a_2(\Lambda, T_\text{t}) + a_4(\Lambda, T_\text{t})\phi^2 + a_6(\Lambda, T_\text{t})\phi^4
    .
\label{EQ_phiM}
\end{equation}
Using again the general expansion introduced in Eq.\ (\ref{EQ_anseries}), one can finally write
    \begin{equation}
    \begin{split}
       &  \text{Critical case: } \\ 
    & \phi\sim (\Lambda-\Lambda_c)^{1/2} \Rightarrow \Delta S \sim \Lambda-\Lambda_c  \Rightarrow \tilde{n} = 1
    \label{EQ_condagammax}        
    \end{split}
    \end{equation}
    
    \begin{equation}
    \begin{split}
        & \text{Tricritical case: } \\
    & \phi\sim (\Lambda-\Lambda_t)^{1/4} \Rightarrow \Delta S \sim (\Lambda-\Lambda_t)^{1/2}  \Rightarrow \tilde{n}_t = \frac{1}{2}
    \label{EQ_condbgammax}        
    \end{split}
    \end{equation}
where $\Lambda_c$ and $\Lambda_t$ are the initial values of $\Lambda$ in the critical and tricritical cases, respectively.
Table \ref{tabI} summarizes the values of the mean-field critical exponents obtained in this section.

\begin{table}[]
    \centering
    \begin{tabular}{c|cccc}
                & $\beta$ &      & $n$ & \\ \hline
        Critical case & $1/2$ & $\phi\sim (T_\text{t}-T)^{1/2}$ & $2/3$ & $\Delta S\sim B^{2/3}$ \\
                & $1$ & $x-x_0\sim (T_\text{t}-T)$ & $1$ & $\Delta S\sim \Lambda$ \\
        Tricritical case & $1/4$ & $\phi\sim (T_\text{t}-T)^{1/4}$ & $2/5$ & $\Delta S\sim B^{2/5}$ \\
                & $1/2$ & $x-x_0\sim (T_\text{t}-T)^{1/2}$ & $1/2$ & $\Delta S\sim \Lambda^{1/2}$
    \end{tabular}
    \caption{Critical exponents obtained using an analysis based on a free energy Landau expansion. We remark that the value given for the tricritical case of $x-x_0\sim (T_\text{t}-T)^{1/2}$ holds for $T<T_\text{t}$, while $x-x_0\sim T_\text{t}-T$ when $T>T_\text{t}$.}
    \label{tabI}
\end{table}

\section{Diluted ferromagnet under pressure}
\label{BEG}

We consider now the explicit case of a diluted ferromagnet containing two types of components. A fraction $1-x$ of magnetic components characterized by a localized magnetic moment that can be oriented up and down, and a fraction $x$ of non-magnetic impurities. The system can be modeled by means of a three state lattice hamiltonian. We thus consider a spin-1 variable $S$ that can take values $S_{\pm}=\pm 1$ when a lattice site is occupied by a magnetic atom in the up and down magnetic states, and $S_0 = 0$ when it is occupied by an impurity. This model can be used to describe the magnetic behaviour of La(Fe$_x$Si$_{1-x}$)$_{13}$ within the ferromagnetic region close to its tricritical point by including magnetovolume effects. The magnetic transition can thus be induced by controlling a mechanical pressure. The presence of a magnetovolume coupling, typically observed in metallic magnetic materials, follows by considering that the magnetic exchange parameter is a function of volume. Therefore, we assume that the hamiltonian of the system is of the form 
\begin{equation}
\begin{split}
{\cal H} = & - J(\omega) \sum_{\langle ij \rangle_{nn}} S_i S_j - H \sum_{i = 1}^N S_i + \mu \sum_{i = 1}^N S_i^2 \\
& +pNv_0\omega + \frac{1}{2}Nv_0K\omega^2,
\end{split}
\end{equation}
where $N$ is the number of lattice sites, $H$ is an external applied magnetic field, $\mu$ a chemical potential that controls the amount of impurities in the system, and $p$ the hydrostatic pressure. Note that application of pressure causes a change of volume
\begin{equation}
\omega=\frac{v-v_0}{v_0},
\label{EQ_omega}
\end{equation}
with an energy cost described by an elastic term associated with a bulk modulus $K$. $v_0$ is the unit cell volume in the absence of external stimuli and in the paramagnetic state, as shown in Eq.\ (\ref{EQ_wmin}) later on. 

The preceding model can be viewed as a generalization of the well known three-states Blume-Emery-Griffiths (BEG) hamiltonian~\cite{Blume1971} where the magnetostructural coupling is introduced as proposed in the Bean-Rodbell (BR) model~\cite{PhysRev.126.104}. We will denote it as the BEG-BR model. It is insighful to remind that the BEG model is a prototype to deal with tricritical points. It was first applied to study $^3$He-$^4$He mixtures where it is known that the $\lambda$-transition line ends at a tricritical point at given amount of fermionic $^3$He atoms. It has been shown that it can be also applied to the study of metamagnetic materials~\cite{Kinkaid1975} as well as of martensitic transformations~\cite{PhysRevB.53.8915}. On the other hand, the BR model was proposed as a generalization of the Ising model that permits describing first-order magnetic transitions. Therefore, the present model is a combination of both the BEG and the BR models, which has the advantage that the transition can be induced by both a primary and a secondary field and shows a dependence on the amount of dilution. From this point of view, we expect that this is a convenient model to study magnetocaloric and barocaloric effects in systems of controllable amount of dilution, and in particular, their behaviour near critical and tricritical points. 

To solve the model we use the variational method and approximate the probability matrix as product of single-density probability matrices ${\boldsymbol{\rho}} = \prod_{i = 1}^N \boldsymbol{\varrho}_i$, where $\boldsymbol{\varrho}_i$ is a $3 \times 3$ matrix with only diagonal terms giving the probability $\varrho_{\alpha}$ that site-$i$ is occupied by an atom with spin $S_{\alpha}$, $\alpha$ standing for $\{+ , -, 0 \}$. The Gibbs free energy density can, therefore, be expressed as
\begin{equation}
\begin{split}
    {\cal G} = & - \frac{1}{2} z J[\text{Tr} \; (\varrho_{\alpha} S_{\alpha})]^2 + \mu [\text{Tr} \; (\varrho_{\alpha} S_{\alpha}^2)] - H [\text{Tr} (\varrho_{\alpha} S_{\alpha})] \\
    & + k_BT [\text{Tr} \; (\varrho_{\alpha} \ln p_{\alpha})]
    +\left(pv_0\omega + \frac{1}{2}v_0K\omega^2\right)[\text{Tr}\; (\varrho_{\alpha})].
    \label{free-energy}    
\end{split}
\end{equation}
In the absence of an applied magnetic field, minimization of ${\cal G}$ under the condition $\text{Tr}\; \varrho_{\alpha} = 1$ leads to
\begin{equation}
\boldsymbol{\varrho} = \frac{1}{{\cal Z}}\left( \begin{matrix}
    e^{-\beta(-zJm + \mu)} & 0 & 0 \\
    0 & e^{-\beta(zJm + \mu)} & 0 \\
    0 & 0 & 1 
\end{matrix} \right),
\end{equation}
where

\begin{equation}
\mathcal{Z}=1+2e^{-\beta \mu}\cosh(\beta zJ m),
\label{EQ_ZBEG}
\end{equation}
and
\begin{equation}
m=\frac{2e^{-\beta \mu}}{\mathcal{Z}}\sinh(\beta z J m)
\label{EQ_mBEG}
\end{equation}
is the primary, ferromagnetic, order parameter. Note that an expression for the secondary parameter $x$ conjugated to the field $\mu$ is
\begin{equation}
1-x=\frac{2e^{-\beta \mu}}{\mathcal{Z}}\cosh(\beta z J m)
.
\label{EQ_xBEG}
\end{equation}

The free energy and entropy functions are, therefore, given as
\begin{equation}
\begin{split}
    \mathcal{G}
= & \frac{1}{2} z J m^2
-k_\text{B}T
\log\left[
1 + 2e^{-\beta \mu}\cosh(\beta z J m)
\right] \\
& -Hm
+pv_0\omega+\frac{1}{2}v_0K\omega^2
,
\label{EQ_FBEG}
\end{split}
\end{equation}
and
\begin{equation}
\begin{split}
    S =k_\text{B}
    \Big( & 
\log\left[1 + 2e^{-\beta \mu}\cosh(\beta z J m) \right] \\
& -\beta z J m^2
+\beta \mu (1-x)
\Big),
\label{EQ_SBEG}
\end{split}
\end{equation}
respectively.

\subsection{Critical temperature and tricritical point driven by $\mu$}
\label{mudriven}

We start by presenting the general qualitative behaviour near the tricritical point obtained in section \ref{critical_exp} by restricting our hamiltonian to the BEG model as a canonical representative. In this situation, $\mu$ plays the role of $\Lambda$ in the absence of an applied pressure, $p=0$, and so there is no need to include a magnetovolume coupling. Hence, we can consider $J$ constant and $K=0$ since these do not couple to the magnetic system.
A relatively lengthy but straightforward Taylor expansion of Eq.\ (\ref{EQ_FBEG}) in powers of $m$ provides the coefficients in Eq.\ (\ref{EQ_Flandau}) of $\mathcal{F}_\text{BEG}$ in this situation, which up to sixth degree are
\begin{equation}
\begin{split}
a_0(\mu,T) & = -k_\text{B}T\log\left(1+2e^{\beta \mu}\right)  \\
a_2(\mu,T) & = zJ\left(1-\frac{zJ}{\eta k_\text{B}T}\right)    \\
a_4(\mu,T) & = \frac{3-\eta}{6\eta^2}\frac{(zJ)^4}{(k_\text{B}T)^3}     \\
a_6(\mu,T) & = \frac{-30+15\eta-\eta^2}{120\eta^3}\frac{(zJ)^6}{(k_\text{B}T)^5}
,
\end{split}
\label{EQ_allaBEG}
\end{equation}
where 
\begin{equation}
\eta=1 + \frac{1}{2}e^{\beta \mu} .
\label{EQ_etaBEG}
\end{equation}
The second-order transition temperature can now be obtained using Eq.\ (\ref{EQ_Tcr})
\begin{equation}
T_\text{c} = \frac{zJ}{k_\text{B}\eta_c}
,
\label{EQ_TtrBEG}
\end{equation}
$\eta_c$ being Eq.\ (\ref{EQ_etaBEG}) for the corresponding transition temperature
\begin{equation}
\eta_c = 1 + \frac{1}{2}\exp\left(\frac{\mu}{k_\text{B}T_\text{c}}\right)
.
\label{EQ_etac}
\end{equation}
The tricritical point at which the phase transition changes its character from second-order to first-order can be found by applying Eq.\ (\ref{EQ_tricritical}). This implies that $\eta_t = 3$, from which, and together with Eqs.\ (\ref{EQ_ZBEG}-\ref{EQ_xBEG}) and (\ref{EQ_allaBEG}-\ref{EQ_TtrBEG}), the transition temperature, $x$, and $\mu$ at the tricritical point can be obtained
\begin{equation}
     T_\text{t} = \frac{zJ}{3k_\text{B}}
     \text{, }\,\,\,\,
     x_\text{t} = \frac{2}{3}
     \text{, }\,\,\,\,
     \mu_\text{t} = \frac{zJ}{3}\log 4
     \text{, }\,\,\,\,
     \eta_\text{t} = 3
     .
\label{EQ_Mtricritical_BEG}
\end{equation}

\subsubsection{Free energy minimization and critical exponents}

Expressions for the critical behaviour within the two cases described in section \ref{critical_exp} can be obtained using previous equations and carrying out some lengthy algebra. We demonstrate and show them in detail for the particular situation of this section in appendix \ref{appendix1}, while in the following the main results obtained after minimizing the Gibbs free energy are discussed.

\begin{figure}[t]
\includegraphics[clip,scale=0.8]{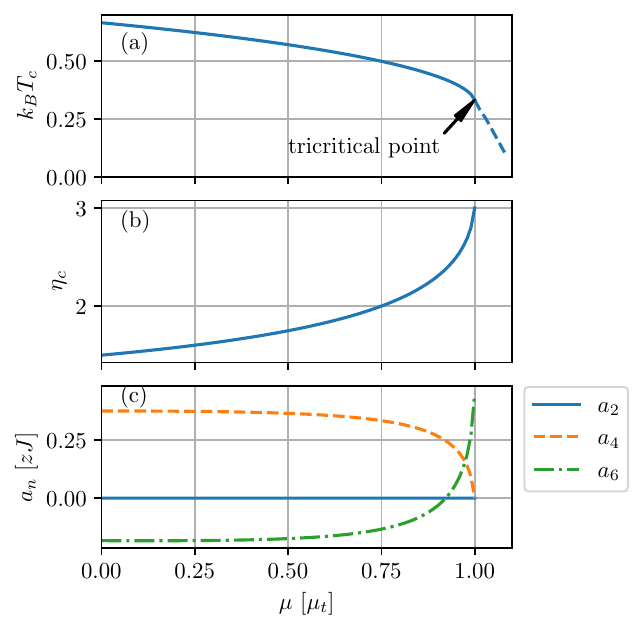}
\caption{Dependence on $\mu$ of the (a) critical transition temperature as well as of (b) $\eta$, and (c) $\{a_n\}$ at $T=T_c$. In panel (a) continuous/discontinuous lines indicate second-/first- order phase transitions, which intersect at the tricritical point. The coefficients $\{a_n\}$ in panel (c) are given in units of $zJ$, while the $\mu$-axis shared by all panels is given in units of $\mu_t=zJ\log(4)/3$, see Eq.\ (\ref{EQ_Mtricritical_BEG}).
}%
\label{Fig1}
\end{figure}

\begin{figure}[t]
\includegraphics[clip,scale=0.8]{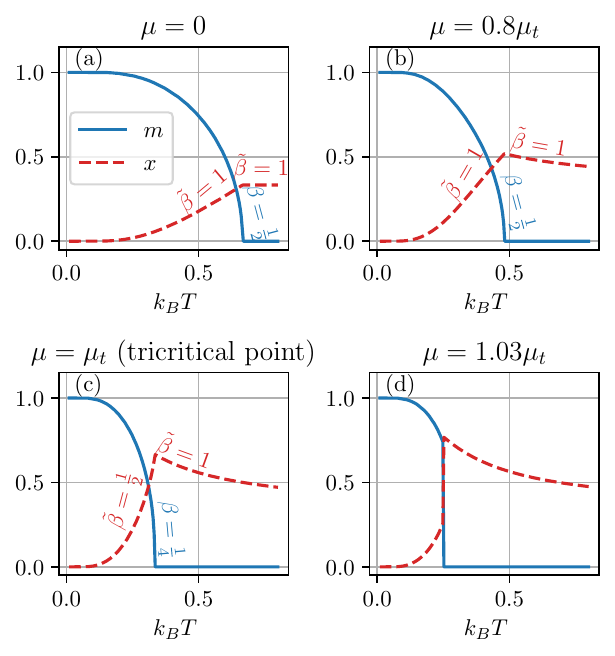}
\caption{Temperature dependence of the primary and secondary order parameters for different values of $\mu$ below (a,b), at (c), and above (d) the tricritical point occurring when $\mu=\mu_t$. The values of the corresponding critical exponents are indicated. Results have been obtained in the absence of both pressure, $p=0$, and magnetovolume coupling, and the model parameters have been normalized to $zJ$.
}%
\label{Fig2}
\end{figure}

Fig.\ \ref{Fig1} shows the dependence of the critical transition temperature, $\eta_c=\eta(\mu, T_c)$, and $a_n(\mu, T_c)$ on the secondary external field $\mu$. For sufficiently large values of $\mu$, $a_6$ changes sign, from negative to positive. This correctly occurs prior approaching the tricritical point, at which $a_4$ becomes negative. Discontinuous phase transitions consequently result beyond this point. The corresponding primary and secondary order parameters as functions of temperature for different values of $\mu$ are shown in Fig.\ \ref{Fig2}. Increasing values of the chemical potential cause a decrease of $T_c$ alongside with an enhancement of $\eta_c$, as observed in Fig.\ \ref{Fig1}(b). This is accompanied by an increment of $x(T_c)$ as given by
\begin{equation}
Nx_c = 1 - \frac{1}{\eta_c},
\label{EQ_x0BEG}
\end{equation}
which can be derived from Eq.\ (\ref{EQ_Nx_x0}) in appendix \ref{appendix1}.
%
\begin{figure}[t]
\includegraphics[clip,scale=0.78]{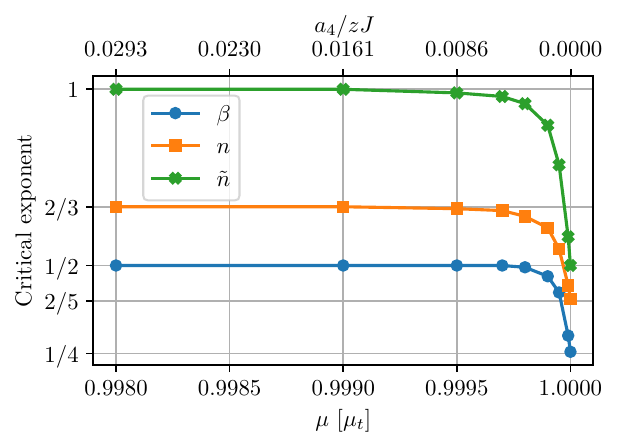}
\caption{Critical exponents right below the tricritical point as functions of the secondary field $\mu$ associated with the change of magnetization ($\beta$) and isothermal entropy change driven by $H$ ($n$) and $\mu$ ($\tilde{n}$) at the critical temperature. Results have been obtained in the absence of both pressure, $p=0$, and magnetovolume coupling.
}%
\label{Fig_critical_exponents_mu}
\end{figure}
%
Indeed, the critical exponent $\beta$ becomes $1/4$ (from $1/2$) at the tricritical point, while the same exponent for $x$, i.e.\ $\tilde{\beta}$, is always 1 above and below the critical temperature expected when $T<T_t$ at the tricritical point, see table \ref{tabI}.

In Fig.\ \ref{Fig_critical_exponents_mu} we plot the dependence on $\mu$ of $\beta$, as well as of the critical exponents associated with the isothermal entropy changes at the critical temperature driven by $H$ and $\mu$, i.e.\ $n$ and $\tilde{n}$, respectively.
The figure shows how the critical exponents change to their tricritical value in a very narrow range of the secondary external field.
We have calculated them numerically by finding the best linear regression of the corresponding thermodynamic change against $\mu^{n}$ using a least squares method. The critical exponent giving the smallest error associated to the regression over a dense range of values of $n$ has been considered as the best estimation.

\begin{figure}[t]
\includegraphics[clip,scale=0.83]{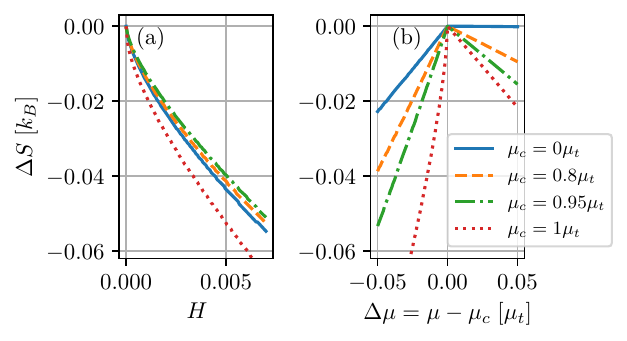}
\caption{Isothermal entropy changes at $T_c$ driven by the application of (a) $H$ and (b) $\mu$. Results are shown for different values of the initial chemical potential ($\mu_c$), which remains constant in panel (b), prior the application of the external field.
The model parameters have been normalized to $zJ$.
}%
\label{Fig_entropy}
\end{figure}

Isothermal entropy changes at the critical transition temperature generated by an applied magnetic field $H$ as well as a change $\Delta\mu = \mu - \mu_c$ are shown in Fig.\ \ref{Fig_entropy}. $\mu_c$, or $\mu_t$ at the tricritical point, are the initial values of the chemical potential in both cases. $\Delta S$ driven by $H$ presents the well-known behaviour with a smaller critical exponent $n_t=2/5 < n=2/3$ at the tricritical point, which yields an enhanced magnetocaloric effect. On the other hand, $\tilde{n}=1$ for both positive and negative values of $\Delta\mu$. The expected decrease of this exponent to $\tilde{n}_t=1/2$ occurs at the tricritical point, as shown in Fig.\ \ref{Fig_critical_exponents_mu}, and for $\Delta\mu<0$. The negative sign of the latter is a consequence of the fact that $m$ remains zero at $T_c$ when the chemical potential is increased instead, as seen in Fig.\ \ref{Fig2}.

\subsection{Effect of pressure}
\label{effect_of_p}

While the study of the effect of $\mu$ has served us to present the general critical and tricritical behaviours within the prototypical BEG model, approaching the tricritical point can be more conveniently achieved experimentally by the application of $p$ instead of changing the chemical potential.
Therefore, in this section we study the case of a diluted ferromagnet under pressure exhibiting a coupling of magnetism with volume, $p$ thus playing the role of $\Lambda$ in this case. The simplest dependence of the magnetic interaction that can be considered and suffices for our purposes is
\begin{equation}
J = J_0 + \alpha \omega,
\label{EQ_MVC}
\end{equation}
where $\alpha$ describes a linear magnetovolume coupling and $J_0$ is the value of the magnetic interaction at $v_0$, see Eq.\ (\ref{EQ_omega}). Minimizing Eq.\ (\ref{EQ_FBEG}) with respect to $\omega$ provides
\begin{equation}
\omega = \frac{1}{v_0 K}\left(\frac{1}{2}z\alpha m^2 - p\right),
\label{EQ_wmin}
\end{equation}
which introduced back to Eq.\ (\ref{EQ_FBEG}) yields
\begin{equation}
\begin{split}
\mathcal{G}
=\,\, &  \frac{1}{2} z J_0 m^2 +\frac{1}{2v_0K}\left[
\frac{3}{4}z^2\alpha^2m^4 - z\alpha p m^2 - p^2
\right]
 \\
& -k_\text{B}T
\log\left[
1 + 2e^{-\beta \mu}\cosh(\beta z [J_0+\frac{1}{2}\alpha\omega] m)
\right] \\
& -Hm
.
\label{EQ_FBEGwmin}    
\end{split}
\end{equation}

As similarly done in section \ref{mudriven}, a Taylor expansion of Eq.\ (\ref{EQ_FBEGwmin}), now in powers of both $m$ and $p$, can be carried out to obtain the Landau coefficients in Eq.\ (\ref{EQ_Flandau}). The most relevant second and fourth order terms are
\begin{equation}
    a_2(p,T) = zJ_0\left(1-\frac{zJ_0}{\eta k_\text{B}T}
\left[1-\frac{\alpha p}{v_0KJ_0}\right]^2
-\frac{\alpha p}{v_0KJ_0}
\right)
,
\label{EQ_a2_BEGp}
\end{equation}
and
\begin{equation}
\begin{split}
   &  a_4(p,T) = 
\frac{z^2}{v_0K}\left[
\frac{3}{2}-2\frac{zJ_0}{\eta k_\text{B}T}
\right]\alpha^2
+\frac{3-\eta}{\eta^2}\frac{z^4J_0^4}{6(k_\text{B}T)^3}
\\
& 
+\frac{3}{2}\frac{z^3}{\eta^2 v_0K k_\text{B}T}\left[
3\frac{\eta}{Kv_0}\alpha^2
-\frac{z^2J_0^3}{(k_\text{B}T)^2}(3-\eta)
\right]\alpha p
 + \mathcal{O}(p^2)
 ,
\end{split}
\label{EQ_a4_BEGp}
\end{equation}
respectively.
Expressions for the zeroth and sixth orders are shown in appendix \ref{appendix2} for completeness.
The second-order critical transition temperature, which follows from $a_2(p,T_c)=0$, is
\begin{equation}
T_c = \frac{zJ_0}{k_\text{B}\eta_c}
\left(
1-\frac{\alpha p}{v_0KJ_0}
\right)
.
\label{EQ_Tcp}
\end{equation}
The value of $T_c$ can be controlled by applying $p$, which decreases or increases it for positive or negative values of $\alpha$, respectively.
Most importantly, both $\alpha$ and $p$ can act as sources to change the sign of $a_4$ to negative and so induce a first-order character of the transition. For $p=0$, Eq.\ (\ref{EQ_a4_BEGp}) reduces at $T=T_c$ to
\begin{equation}
    a_4(p=0,T_c) = zJ_0\left[
\frac{3-\eta_c}{6\eta_c^2}
-\frac{z\alpha^2}{2J_0v_0K}
\right]
 .
\label{EQ_a4_BEGp0}
\end{equation}
The previous result shows that $a_4<0$ for sufficiently large values of the magnetovolume coupling $\alpha$ independently of its sign.
On the other hand, the effect of $p$ on $a_4$ is rather complex. Using Eq.\ (\ref{EQ_a4_BEGp0}) one can see that the first term in the second line of Eq.\ (\ref{EQ_a4_BEGp}) is negative for small values of $p$. Consequently, when $\alpha>0$ the application of pressure $p>0$ contributes to a negative sign of $a_4$ and so it induces a first-order character for the transition. However, the opposite trend occurs if $\alpha<0$.

\section{Comparison with experiments}
\label{experiments}

\subsection{Numerical calculation of critical exponents}
\label{numerical_n}

The interest of this section is to present a comparison of critical exponents for the entropy change with experiments near or at tricritical points. However, as explained in the introduction, measurements of caloric effects around tricritical points are rare, and most of the few of them that exist are not extensive enough to perform a successful analysis. To the best of our knowledge, available experimental data for La(Fe$_{x}$Si$_{1-x}$)$_{13}$ and MnSi compounds is sufficient to carry out a reasonable comparison, which we explain and discuss in Sects.\ \ref{LaFeSi_comp} and \ref{MnSi_comp}, respectively.

The corresponding reported measurements in these two materials that we can use have been made under the presence of an applied magnetic field for the initial conditions. This calls for careful considerations since such a situation is slightly different from the one in the theoretical developments presented in section \ref{critical_exp} where, for the sake of simplicity, it has been assumed that the line of critical points occurs for a zero primary field. To address this issue, we estimate the critical exponent by subtracting the initial critical (or tricritical) state. In the case of an applied pressure $p$ under a constant critical magnetic field (barocaloric effect) we have
\begin{equation}
\Delta S = S(H_c,p) - S(H_c,0) \propto p^{n}
,
\label{EQ_subtractBCE}
\end{equation}
while for an applied magnetic field (magnetocaloric effect) without pressure
\begin{equation}
\Delta S = S(H,0) - S(H_c,0) \propto (H-H_c)^{n}
,
\label{EQ_subtractMCE}
\end{equation}
where $H_c$ is the critical magnetic field, tricritical at $T_t$. We then perform a series of linear regressions of Eqs.\ (\ref{EQ_subtractBCE}) or (\ref{EQ_subtractMCE}) using a least squares method for a dense range of values of $n$, as similarly done to obtain Fig.\ \ref{Fig_critical_exponents_mu}. The value with the smallest error associated with the linear regression is considered as the best estimation.

\subsection{Application to La(Fe$_{x}$Si$_{1-x}$)$_{13}$}
\label{LaFeSi_comp}

The general aspects of the first-order paramagnetic-ferromagnetic phase transition and related tricritical point observed in the La(Fe$_{x}$Si$_{1-x}$)$_{13}$ magnetocaloric compound~\cite{PhysRevB.65.014410, PhysRevB.67.104416} can be described well by the BEG-BR model both qualitatively and quantitatively. The fundamental origin of its first-order character is a large magnetovolume coupling, as recently shown and quantified fully from first principles calculations~\cite{Mendive_Tapia_2023}. Such a transition is consequently accompanied by a substantial spontaneous volume change of approximately $1\%$~\cite{PhysRevB.65.014410}. The material's volume is larger right below $T_c$ within the low temperature ferromagnetic state, which corresponds to the so-called negative thermal expansion.
Hence, to model La(Fe$_{x}$Si$_{1-x}$)$_{13}$ we must set a sufficiently large and positive value of the magnetovolume constant, as seen in Eq.\ (\ref{EQ_wmin}). We have found that $\alpha=7 zJ_0$ together with $\mu=0.95\mu_t$ correctly establish a first-order transition that can share quantitative features observed in this material, particularly for La(Fe$_{0.86}$Si$_{0.14}$)$_{13}$ ($x=0.86$). For example, the product of the bulk modulus with $v_0$ can be adjusted to $Kv_0=100 zJ_0$ in order to render the right order of magnitude of the experimental spontaneous volume change at the transition, that is $\Delta\omega\approx 1\%$.
The chosen value of $\mu$ does not have important implications as soon as it is smaller than $\mu_t$ and so it is not the cause of the first-order nature.

\begin{figure}[t]
\includegraphics[clip,scale=0.8]{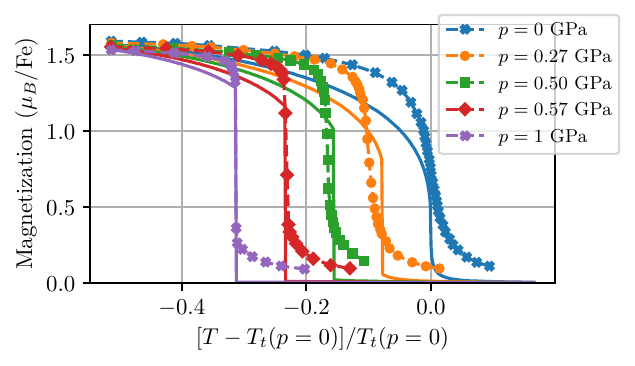}
\caption{Comparison of theory (continuous lines) with experimental measurements of La(Fe$_{0.86}$Si$_{0.14}$)$_{13}$'s magnetization extracted from reference \cite{PhysRevB.65.014410} (data points) for increasing values of $p$. The model parameters used, normalized to $zJ_0$, have been $\alpha=7$, $\mu=0.95\mu_t$, and $Kv_0=100$, under the application of a magnetic field $H_t^\text{theo}=2.54\cdot 10^{-4}$ to reach the tricritical point occurring at $k_\text{B}T_t^\text{theo}=0.403$.
The corresponding values of $p$ applied in the model range as $p=0\rightarrow 1.7$ linearly to match experiment, and the order parameter $m$ has been multiplied by the magnetic moment magnitude of Fe.
The experimental value of $T_t$ has been chosen as the temperature where the magnetization profile presents an inflection point in the absence of pressure.
}%
\label{Fig_LaFeSi}
\end{figure}

Fujita \textit{et al}.\ have shown experimentally that the application of a relatively small tricritical magnetic field of $H_t^\text{exp}=0.3$T to La(Fe$_{0.86}$Si$_{0.14}$)$_{13}$ removes most of its first-order character~\cite{PhysRevB.65.014410}. As suggested by Eq.\ (\ref{EQ_a4_BEGp}), the material's positive magnetovolume coupling ($\alpha>0$) implies that the magnetic discontinuity can be recovered in this scenario by subjecting the material to an external pressure. In other words, a tricritical point is effectively reached at $H_t^\text{exp}
$ and left away towards the region of first-order transitions by subsequently applying $p$. We first study such a tricritical point by numerically minimizing the Gibbs free energy with La(Fe$_{0.86}$Si$_{0.14}$)$_{13}$'s parameters mentioned above.

Fig.\ \ref{Fig_LaFeSi} shows the corresponding results computed for the temperature-dependent magnetization alongside with the experimental measurements~\cite{PhysRevB.65.014410}. Theoretical values of the applied magnetic field, which remains constant, and of hydrostatic pressures have been chosen to suitably reach a tricritical point at $p=0$ and decrease the critical temperature proportionally as in experiments, respectively. Detail in these values is given in the caption of the figure.
In particular, a tricritical temperature of $T_t^\text{exp}=210$K can be established in the calculation by setting $zJ_0 = 54.3$meV, see Eq.\ (\ref{EQ_Mtricritical_BEG}). This value gives $H_t^\text{theo}=0.16$T, which is in reasonable agreement with experiment.
Indeed, the calculations simulate experimental trends very well and satisfactorily describe the recovery of the first-order character caused by $p$ near the tricritical point.

We can now proceed to calculate the critical exponent associated with the entropy change driven by the application of $p$.
The available experimental data in reference \cite{PhysRevB.65.014410} reports the pressure-dependence of the square of the spontaneous magnetization change at the transition temperature, which should be approximately proportional to $\Delta S$ [see Eq.\ (\ref{EQ_DSphi})]. 
We highlight, however, that the exponent associated with this entropy change does not correspond exactly to $\tilde{n}$ in table \ref{tabI} since it is not isothermal. We thus name it $\tilde{n}'$ to avoid confusion. Nevertheless, it can still be used to study the thermodynamic response of the material and its relation to an applied field when compared with the model's outputs. To this end, we calculate the exact same theoretical tricritical exponent, which can be directly obtained by accessing to the corresponding squares of the computed spontaneous magnetization changes in Fig.\ \ref{Fig_LaFeSi}.
The obtained results are $\tilde{n}'_\text{theo.}=0.77$ and $\tilde{n}'_\text{exp.}=0.92$, estimated using Eq.\ (\ref{EQ_subtractBCE}) as described in \ref{numerical_n} and with least squares residuals smaller than $10^{-3} \mu_\text{B}^2$.
The agreement between theory and experiment is satisfactory and corroborates that $p$ is a secondary field.
A possible cause of the observed discrepancy might be the fact that $H_t^\text{exp}=0.3$T could potentially be an overestimation, which should artificially raise the value of $\tilde{n}'$.

\subsection{MnSi}
\label{MnSi_comp}

\begin{figure}[t]
\includegraphics[clip,scale=0.81]{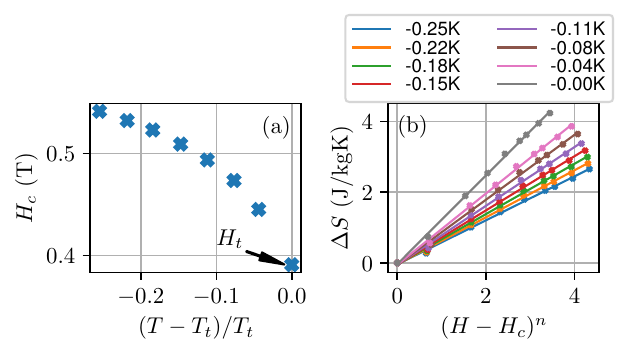}
\includegraphics[clip,scale=0.81]{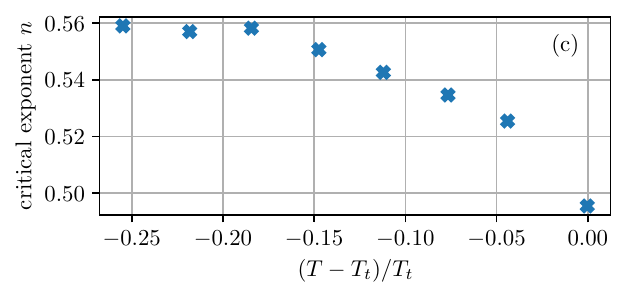}
\caption{
(a) Temperature dependence of the critical magnetic field of MnSi. The tricritical value at $T_t$ is indicated.
(b) Linear regression of Eq.\ (\ref{EQ_subtractMCE}) with the smallest associated error at different relative temperatures $\frac{T-T_t}{T_t}$.
(c) Experimental critical exponent for the isothermal entropy change at the critical temperature, $n$.
The used experimental data for MnSi has been extracted from Refs~\cite{Samatham2017,doi:10.1126/science.1166767}.
}
\label{Fig_beta_p}
\end{figure}

The temperature-magnetic field phase diagram of MnSi presents a tricritical point at $T_t=28.5$K under a value of $H_t=0.340$T~\cite{Bauer2013, doi:10.1126/science.1166767}. Entropy measurements have been reported around such a tricritical point~\cite{Samatham2017}, which can be used to estimate the associated critical exponent $n$.
To this end, we have first performed an additional linear regression of the experimental data in order to extrapolate the dependence on temperature of the critical magnetic field in MnSi's diagram, which we show in Fig.\ \ref{Fig_beta_p}(a).
These values have then been used to calculate the initial entropy and subtract it in order to adequately obtain $n$ at different temperatures at and below $T_t$ by applying Eq.\ (\ref{EQ_subtractMCE}), as explained in section \ref{numerical_n}.
The corresponding final linear regressions and values of the critical exponents are plotted in panels (b) and (c), respectively. An inspection of the latter shows that $n$ is a little bit lower than $1/2$ at $T_t$ (the mean-field value being 0.4) and slightly increases until it becomes approximately constant and equal to 0.56 (the mean-field value being 2/3) at lower temperatures. These values are duly smaller than the unity, which allows us to identify and corroborate that $H$ is the principal conjugated field of the system in accordance to table \ref{tabI} and in comparison with Fig.\ \ref{Fig_critical_exponents_mu}. Our findings are in good agreement with the reported value of $n=0.51$ estimated using the same experimental data and similar procedure, taking instead the largest value of the magnetocaloric effects at each applied field~\cite{Samatham2017}.

\section{Conclusions}
\label{conclusions}

In summary, we have first presented the general features of caloric effects in materials near and at tricritical points by means of a Landau expansion. We then constructed and solved within a mean-field approximation a broadly applicable Hamiltonian based on the combination of the paradigmatic Blume-Emery-Griffiths (BEG) and Bean-Rodbell (BR) models. Critical and tricritical exponents driven by both primary and secondary conjugated fields have been determined, as well as the crossover induced by the latter. To the best of our knowledge, this is the first time that the value and behaviour of such fundamental tricritical properties have been reported.

Both the Landau expansion and the extended BEG-BR model are powerful tools for analyzing the role and caloric response of applied fields observed experimentally in tricritical materials. Our developments are also useful for investigating the sequential and simultaneous application of multiple fields, which lays out the groundwork for future research that we plan to carry out focused on multicaloric effects within this BEG-BR model.

Tricritical points fundamentally represent one of the most important research areas towards the optimization of caloric effects, crucial for their success as competitive ecological technology. However, many of the materials exhibiting tricritical points present not deep enough investigations regarding their caloric behaviour or directly lack of them.
Our best efforts have corroborated and quantified the roles of applied fields by estimating their tricritical exponents in only two materials, the well-known La(Fe$_{x}$Si$_{1-x}$)$_{13}$ and MnSi compounds.
New experimental research on caloric effects of current and yet-to-be-discovered materials with functional tricritical points is thus needed and deserves being made, which could be readily analysed using our theoretical developments.

\begin{acknowledgements}
We acknowledge support from MCIN / AEI / 10.13039 / 501100011033
(Spain) under Grant No.\ PID2020-113549RB-I00/AEI. E.\ M.-T.\ is grateful for funding received from the European Union's Horizon 2020 research and innovation programme under the Marie Sklodowska-Curie grant agreement No.\ 101025767.
\end{acknowledgements}

\appendix
\section{Detailed expressions in the critical limits of the BEG-BR model}
\label{appendixA}

In this appendix we provide detailed expressions regarding and derived from the Landau coefficients in both the critical and tricritical limits of the BEG-BR model presented in section \ref{BEG}. Pertinent results obtained for the application of $\mu$ and $H$ (Sect.\ \ref{mudriven}) or $p$ (Sect.\ \ref{effect_of_p}) are given in \ref{appendix1} and \ref{appendix2}, respectively. We remark that $m$, $H$, and $\mu$ (or $p$) correspondingly play the role of $\phi$, $B$, and $\Lambda$ in section \ref{critical_exp}.

\subsection{Application of $H$ and $\mu$}
\label{appendix1}

In the critical case $T\rightarrow T_\text{c}$, $a_2(\mu,T)$, Eq.\ (\ref{EQ_allaBEG}) and Eq.\ (\ref{EQ_etaBEG}) become
\begin{equation}
  a_2(\mu,T)\rightarrow -zJ \frac{T_\text{c}-T}{T_\text{c}}
\label{EQ_a2BEGlim}
\end{equation}
and
\begin{equation}
\begin{split}
   \eta\rightarrow & 1+e^{\frac{\mu}{k_\text{B}T_\text{c}}}\left[
\frac{1}{2} + \frac{\mu}{k_\text{B}T_\text{c}} \frac{T_\text{c}-T}{T_\text{c}}
\right] \\
& = \eta_c + e^{\frac{\mu}{k_\text{B}T_\text{c}}}\frac{\mu}{k_\text{B}T_\text{c}} \frac{T_\text{c}-T}{T_\text{c}}
,
\label{EQ_etaBEGlim}
\end{split}
\end{equation}
respectively.
The same limit for the first order derivatives of the free energy coefficients with respect to $\mu$ in Eq.\ (\ref{EQ_allaBEG}) directly follow
\begin{equation}
\begin{split}
  &  \frac{\partial a_0}{\partial \mu} = \frac{1}{1+\frac{1}{2}e^{\beta \mu}} = \frac{1}{\eta} \\
   & \Rightarrow
    \lim_{T\rightarrow T_\text{c}}\frac{\partial a_0}{\partial \mu}
    = \frac{1}{\eta_c} 
    -e^{\frac{\mu}{k_\text{B}T_\text{c}}}\frac{\mu}{2k_\text{B}T_\text{c}\eta_c^2} \frac{T_\text{c}-T}{T_\text{c}} \\
    & +\mathcal{O}\left(\left[\frac{T_\text{c}-T}{T_\text{c}}\right]^2\right)
    ,
\label{EQ_a2derivM_BEG}    
\end{split}
\end{equation}
and
\begin{equation}
\begin{split}
    & \frac{\partial a_2}{\partial \mu} = \frac{(zJ)^2}{(k_\text{B}T)^2}\frac{e^{\beta \mu}}{2\eta^2} \\
    & \Rightarrow
    \lim_{T\rightarrow T_\text{c}}\frac{\partial a_2}{\partial \mu}
    = \frac{e^{\frac{\mu}{k_\text{B}T_\text{c}}}}{2}
    +\mathcal{O}\left(\left[\frac{T_\text{c}-T}{T_\text{c}}\right]\right)
    ,
\label{EQ_Nx_x0}
\end{split}
\end{equation}
which allow to write Eq.\ (\ref{EQ_xbeta}) in the model:
\begin{equation}
N(x-x_c) =
e^{\frac{\mu}{k_\text{B}T_\text{c}}}\frac{\mu}{2k_\text{B}T_\text{c}\eta_c^2} \frac{T_\text{c}-T}{T_\text{c}}
-\frac{1}{4}e^{\frac{\mu}{k_\text{B}T_\text{c}}} m^2
,
\label{EQ_xlimBEG}
\end{equation}
where $x_c$ is defined in Eq.\ (\ref{EQ_x0BEG}).
Note that the first term in the right hand side of this equation arises from the condition given in Eq.\ (\ref{EQ_xBEG}).

On the other hand, the expansion of the entropy in Eq.\ (\ref{EQ_SBEG}) reads as
\begin{equation}
S_\text{BEG} = S_\text{BEG,0} +\frac{1}{2} S_\text{BEG,2} m^2 
+\mathcal{O}(m^4)
,
\label{EQ_SBEG_exp}
\end{equation}
the first two non-zero coefficients being 
\begin{equation}
S_\text{BEG,0} = 
k_\text{B}\left[
\log\left(1+2e^{-\beta \mu}\right)
+\frac{\mu}{zJ}\right]
\label{EQ_SBEG_exp0}
,
\end{equation}
and
\begin{equation}
S_\text{BEG,2} = 
\frac{1}{T_c}\left[
(\eta_c-1)\mu
-zJ
\right]
,
\label{EQ_SBEG_exp2}
\end{equation}
at the critical temperature.
Hence, the isothermal entropy change can be approximated as
\begin{equation}
\Delta S_\text{BEG}
\approx 
\frac{1}{2}S_\text{BEG,2} m^2
= \frac{1}{2T_c}\left[
(\eta_c-1)\mu
-zJ
\right]m^2
.
\label{EQ_DeltaS_BEG}
\end{equation}
In order to study the isothermal entropy change generated by applying $\mu$ from a given initial value $\mu_c$ at $H=0$, i.e.\ $\Delta \mu = \mu-\mu_c$, it is useful to write Eq.\ (\ref{EQ_derphi}) setting $T=T_\text{c}$. To this end, one should find the corresponding free energy coefficients under these conditions, which write as
\begin{equation}
a_2 = zJ\left[
1-\frac{\eta_c}{1+(\eta_c - 1)e^{\frac{\Delta \mu}{k_\text{B}T_\text{c}}}}
\right]
,
\label{EQ_BEGcoeffsa2}
\end{equation}
\begin{equation}
a_4 = zJ\eta_c^3
\frac{2-(\eta_c - 1)e^{\frac{\Delta \mu}{k_\text{B}T_\text{c}}}}{6\left(
1+(\eta_c - 1)e^{\frac{\Delta \mu}{k_\text{B}T_\text{c}}}
\right)^2}
,
\label{EQ_BEGcoeffsa4}
\end{equation}
and 
\begin{equation}
\begin{split}
& a_6 = zJ\eta_c^5 \times \\
& \frac{-30+15\left(1+(\eta_c - 1)e^{\frac{\Delta \mu}{k_\text{B}T_\text{c}}}\right)
-\left(1+(\eta_c - 1)e^{\frac{\Delta \mu}{k_\text{B}T_\text{c}}}\right)^2}{120\left(
1+(\eta_c - 1)e^{\frac{\Delta \mu}{k_\text{B}T_\text{c}}}
\right)^3}
.
\label{EQ_BEGcoeffsa6}
\end{split}
\end{equation}
Their expansions in powers of $\Delta \mu$ up to second degree are
\begin{equation}
a_2 = zJ\left[
\frac{\eta_c-1}{\eta_c}\beta_c\Delta \mu
-\frac{\eta_c^2-3\eta_c+2}{2\eta_c^2}\beta_c^2\Delta \mu^2
\right]
,
\label{EQ_BEGcoeffsa2M}
\end{equation}
\begin{equation}
\begin{split}
a_4 =zJ\eta_c^2\Big[ & 
\frac{3-\eta_c}{6\eta_c^2}
+\frac{(\eta_c-6)(\eta_c-1)}{6\eta_c^3}\beta_c\Delta \mu \\
& -\frac{(\eta_c-1)(\eta_c^2-14\eta_c+18)}{12\eta_c^4}\beta_c^2\Delta \mu^2
\Big]
,
\label{EQ_BEGcoeffsa4M}
\end{split}
\end{equation}
and
\begin{equation}
\begin{split}
a_6 = \,\, &  zJ\eta_c^5\Big[
\frac{-30+15\eta_c-\eta_c^2}{120\eta_c^3} \\
& +\frac{\eta_c^3-31\eta_c^2+120\eta_c-90}{120\eta_c^4}\beta_c\Delta \mu \\
& -\frac{\eta_c^4-63\eta_c^3+422\eta_c^2-720\eta_c+360}{240\eta_c^5}\beta_c^2\Delta \mu^2
\Big]
,
\end{split}
\label{EQ_BEGcoeffsa4M}
\end{equation}
where $\beta_c=\frac{1}{k_\text{B}T_\text{c}}$. Therefore,
\begin{equation}
\begin{split}
0 = \,\, & 
\frac{\eta_c-1}{\eta_c}\beta_c\Delta \mu
+\frac{3-\eta_c}{6}m^2 \\
& 
+\frac{-30+15\eta_c-\eta_c^2}{120}\eta_c^2m^4+\mathcal{O}(m^6)+\mathcal{O}(\Delta \mu m^2)
.
\label{EQ_DeltaMBEG}
\end{split}
\end{equation}

From these results, we can finally write expressions for the critical [$\lim_{T\rightarrow T_\text{t}^{-}} a_2(\mu, T) \ll \lim_{T\rightarrow T_\text{t}^{-}} a_4(\mu, T)$] and tricritical [$\lim_{T\rightarrow T_\text{t}^{-}} a_2(\mu, T) \gg \lim_{T\rightarrow T_\text{t}^{-}} a_4(\mu, T)$] cases, as given in the following.

\subsubsection{Critical case}

From Eq.\ (\ref{EQ_conda}) we directly find the temperature-dependence the primary order parameter in the critical case
\begin{equation}
    m = \left[\frac{6}{(3-\eta_c)\eta_c} \frac{T_\text{t}-T}{T_\text{t}}
    \right]^{1/2}
    ,
\label{EQ_mBEGbetaA}
\end{equation}
This result together with Eq.\ (\ref{EQ_xlimBEG}) can be used to write expressions for the secondary order parameter,
\begin{equation}
    N(x-x_c) = 
    \left(
    e^{\frac{\mu}{k_\text{B}T_\text{t}}}\frac{\mu}{2k_\text{B}T_\text{t}\eta_c^2} 
    - \frac{6}{(3-\eta_c)\eta_c}
    \right)
    \frac{T_\text{t}-T}{T_\text{t}} 
\label{EQ_xBEGbetaA}
\end{equation}
if $T<T_\text{t}$, and
\begin{equation}
    N(x-x_c) =
    e^{\frac{\mu}{k_\text{B}T_\text{t}}}\frac{\mu}{k_\text{B}T_\text{t}\eta_c^2} \frac{T_\text{t}-T}{T_\text{t}}
\label{EQ_xBEGbetaAs}
\end{equation}
if $T>T_\text{t}$.
On the other hand, Eqs.\ (\ref{EQ_condagamma}) and (\ref{EQ_DeltaMBEG}) result in
\begin{equation}
m = \left[
\frac{1}{3-\eta_c}\frac{1}{zJ\eta_c}H
\right]^{1/3}
,
\label{EQ_mHBEGa}
\end{equation}
and
\begin{equation}
m = \left[
-\frac{6(\eta_c-1)}{(3-\eta_c)\eta_c}\beta_c\Delta \mu
\right]^{1/2}
,
\label{EQ_mMBEGa}
\end{equation}
which from Eq.\ (\ref{EQ_DeltaS_BEG}) finally provide expressions for the isothermal entropy changes
\begin{equation}
\Delta S_\text{BEG} \approx
\frac{(\eta_c-1)\mu
-zJ}{2T_c}
    \left[
\frac{1}{3-\eta_c}\frac{1}{zJ\eta_c}H
\right]^{2/3}
,
\label{EQ_DeltaSHBEGa}
\end{equation}
and
\begin{equation}
\Delta S_\text{BEG} \approx
\frac{(\eta_c-1)\mu
-zJ}{T_c}
\frac{3(1-\eta_c)}{\eta_c(3-\eta_c)}\beta_c\Delta \mu
,
\label{EQ_DeltaSMBEGa}
\end{equation}
respectively.

\subsubsection{Tricritical case:}

The primary order parameter in the tricritical case can be directly obtained using now Eq.\ (\ref{EQ_condb}),
\begin{equation}
    m = \left[\frac{120}{(-30+15\eta_c-\eta_c^2)\eta_c^2} \frac{T_\text{t}-T}{T_\text{t}}
    \right]^{1/4}
    ,
\label{EQ_mBEGbetaB}
\end{equation}
which together with Eq.\ (\ref{EQ_xlimBEG}) yields
\begin{equation}
    N(x-x_0) = 
    -\frac{1}{4}
    \left[\frac{120}{(-30+15\eta_c-\eta_c^2)\eta_c^2} \frac{T_\text{t}-T}{T_\text{t}}
    \right]^{1/2} 
\label{EQ_xBEGbetaB}
\end{equation}
if $T<T_\text{t}$, and
\begin{equation}
    N(x-x_0) = 
    e^{\frac{\mu}{k_\text{B}T_\text{t}}}\frac{\mu}{k_\text{B}T_\text{t}\eta_c^2} \frac{T_\text{t}-T}{T_\text{t}}
\label{EQ_xBEGbetaBs}
\end{equation}
if $T>T_\text{t}$. As similarly done for the critical case, Eq.\ (\ref{EQ_condbgamma}) and (\ref{EQ_DeltaMBEG}) can be used to write
\begin{equation}
m = \left[
\frac{120}{-30+15\eta_c-\eta_c^2}\frac{1}{zJ\eta_c^2}H
\right]^{1/5}
,
\label{EQ_mHBEGb}
\end{equation}
and
\begin{equation}
m = \left[
-\frac{40}{27}\beta_c\Delta \mu
\right]^{1/4}
,
\label{EQ_mHBEGb}
\end{equation}
where $\eta=\eta_\text{t}=3$ has been taken. These expressions together with Eq.\ (\ref{EQ_DeltaS_BEG}) can be used to finally obtain the corresponding isothermal entropy changes,
\begin{equation}
\Delta S_\text{BEG} \approx 
\frac{(\eta_c-1)\mu
-zJ}{T_c}
\left[
\frac{120}{-30+15\eta_c-\eta_c^2}\frac{1}{zJ\eta_c^2}H
\right]^{2/5}
\label{EQ_DeltaSHBEGb}
\end{equation}
and

\begin{equation}
\Delta S_\text{BEG} \approx
\frac{(\eta_c-1)\mu
-zJ}{T_c}
\left[
-\frac{40}{27}\beta_c\Delta \mu
\right]^{1/2}
.
\label{EQ_DeltaSMBEGb}
\end{equation}

\subsection{Application of $p$}
\label{appendix2}

As explained in section \ref{effect_of_p}, carrying out a Taylor expansion of Eq.\ (\ref{EQ_FBEGwmin}) in powers of $m$ and $p$ directly provides the coefficients in Eq.\ (\ref{EQ_Flandau}). The remaining zeroth and sixth order coefficients are
\begin{equation}
 a_0(p,T)= -k_\text{B}T\ln(1+2e^{-\beta \mu}) - \frac{p^2}{2v_0K}
 ,
\label{EQ_allaBEGp_0}
\end{equation}
and 
\begin{equation}
\begin{split}
 a_6(p,T)= \,\, & \frac{z^4}{120\eta k_\text{B}T}
\Bigg[
-30\alpha^2\left(\frac{3\alpha^2}{(v_0K)^2} 
+ \frac{2zJ_0^3}{v_0K(k_\text{B}T)^2} \right) \\
& -\frac{z^2J_0^6}{(k_\text{B}T)^4}
+\frac{15}{\eta}\frac{zJ_0^3}{(k_\text{B}T)^2}\left( \frac{12\alpha^2}{v_0K} + \frac{zJ_0^3}{(k_\text{B}T)^2} \right) \\ 
 & -\frac{30}{\eta^2}\frac{z^2J_0^6}{(k_\text{B}T)^4}
\Bigg] 
 + \mathcal{O}(p)
 .
\end{split}
\label{EQ_allaBEGp_6}
\end{equation}

\bibliography{./bibliography.bib}
\end{document}